\documentclass[twocolumn]{jpsj2} 
\title{Physics at International Linear Collider (ILC)}

\author{Hitoshi Yamamoto$^{1}$\thanks{E-mail address: yhitoshi@awa.tohoku.ac.jp,
Present address: Graduate School of Science, Tohoku University, Sendai.}
}

\inst{$^{1}$Tohoku University\\
}

\abst{International Linear Collider (ILC) is an electron-positron collider 
with the initial center-of-mass energy of 500 GeV which is upgradable to 
about 1 TeV later on.  Its goal is to study the
physics at TeV scale with unprecedented high sensitivities. 
The main topics include precision measurements of the Higgs particle properties,  
studies of supersymmtric particles and the underlying theoretical structure 
if supersymmetry turns out to be realized in nature, probing 
alternative possibilities for the origin of mass, and the cosmological 
connections thereof. In many channels, Higgs and leptonic sector in 
particular, ILC is substantially more sensitive than LHC, and is complementary
to LHC overall. In this short article, we will have a quick look at the 
capabilities of ILC.
}

\kword{ILC, Higgs boson, New Physics, supersymmetry, extra dimensions, cosmology, LHC}

\begin{document}
\maketitle

\section{Introduction} 

The standard model is an astonishingly successful theory in describing 
what have been observed in the field of elementary particles. The Higgs particle, which
gives mass to all massive particles, is at the
core of the standard model, but so far has not been found. Furthermore, if one tries to
calculate the radiative correction to the mass squared of Higgs, it diverges 
quadratically with the cut off energy, and if one assumes that the standard model is correct 
up to the energy scale of the grand unification ($\sim 10^{16}$ GeV), 
the correction to the Higgs mass becomes 
the order of the grand unification scale itself. Since precision measurements so far
shows that the standard model Higgs should lie below $\sim$200 GeV, this is only possible if the
original mass and the correction are canceling out to an astonishing precision. This unpleasant
situation is referred to as the fine-tuning problem, or the naturalness problem. The problem
is in part caused by the large difference in energy scale from the Higgs mass to the
grand unification scale, and in this context, it is refered to as
the hierarchy problem\cite{hierarchy}. 
Also, the standard model does not include the gravitational force.

A theoretical solution to the fine-tuning problem is provided by supersymmetry 
(SUSY)\cite{SUSY} which postulates that 
every particle in the standard model has its so-called superpartner (called a super
particle or a s-particle) whose spin differs by one half from that of the original particle. 
Not only the inclusion of the super particles naturally
cancels out the quadratic divergence of the Higgs mass correction, in SUSY the gauge
coupling constants converges to a single value at the grand unification scale. Furthermore,
the gravitational force can also be naturally incorporated in SUSY. As a bonus, SUSY has 
candidates for the dark matter which is thought to consist of unknown stable massive particles
and accounts for one quarter of the energy of the universe. 

Even though SUSY is an attractive theory with many merits for us, nature of course
would not care about our conveniences. There are several alternative models that address
the fine-tuning problem, and some of them may have connection to the reality of nature.
Examples are the models with extra dimensions which postulate the existence of 
space dimensions more than our 3(space)+1(time) dimensions\cite{extradim}, 
and the little Higgs model\cite{littleH}
where the Higgs particle is considered to be composite. 

The physics potential of ILC has been extensively studied and documented.\cite{DCR,
TeslaTDR,JLCphys,GLCroadmap,Snowmass01}
As we will see below, the standard model Higgs particle will have distinctive signals at ILC, and
SUSY and other alternative models also have many possibilities of being found and studied at ILC.
The advantage of ILC with respect to LHC is in the general cleanliness of the events where
two elementary particles (an electron and a positron) with known kinematics and spin define
the initial state, and the high resolutions of the detector that are made possible by the
relatively low absolute rate of background events. The capability of ILC is further 
enhanced by the options such as the $\gamma\gamma$ collision,
$e^-e^-$ collision, and $Z$-pole running (`Giga-Z').

\section{ILC machine parameters and detectors}

The basic parameters, such as energy and luminosity, of ILC are described in the
parameter report\cite{parameter}. The baseline machine allows for a center-of-mass energy
range between 200 GeV and 500 GeV and luminosity of 500 fb$^{-1}$
in the first four years of running not counting the year zero. The energy scan
is possible at any energy within the range, and the electron polarization
is at least 80\%. Two detectors are expected which may be in a push-pull
configuration. 

For each of the two beams, a bunch is $\sigma_y=$ 5.7 nm high, 
$\sigma_x=$ 655 nm wide and $\sigma_z=$ 300 $\mu$m long, 
and contains $2\times 10^{10}$ particles. About 3000 bunches
with 308 ns bunch separtion form a train of about 1ms which 
comes with 5 Hz repetition rate. The collision occurs with
crossing angle of 14 mrad.

The highest priority beyond the baseline is the energy upgrade to approximately
1 TeV, and the upgraded machine should be able to collect 1 ab$^{-1}$ in 
3 to 4 years after the baseline running. The options include: running at 
500 GeV to double the luminosity to 1 ab$^{-1}$, $e^-e^-$ collision, positron
polarization of 50\% or more, $Z$-pole running, $WW$ threshold running, and 
$\gamma\gamma$ and $e^-\gamma$ collisions using backscattered laser beams. 
The priorities of these options will depend on the results of LHC and the baseline
ILC.\cite{LHC-ILC} In the following, the baseline machine with 200 to 500 GeV
center-of-mass energy is assumed unless stated otherwise. 

The physics of ILC is realized through synthesis of unprecedented 
performances of both machine and detectors. ILC detectors can take
advantage of the relatively low rates and low radiation doses to 
achieve momentum resolution that is order of magnitude better,
jet energy resolution factor of two better, and the vertex resolution
several times better than those at the previous electron-positron
colliders. As we will see below, these performances are not
overkill; rather, they are needed to realize the physics potential of ILC.

\section{Standard model particles}

\begin{figure}[tb]
\begin{center}
\includegraphics[width=3in]{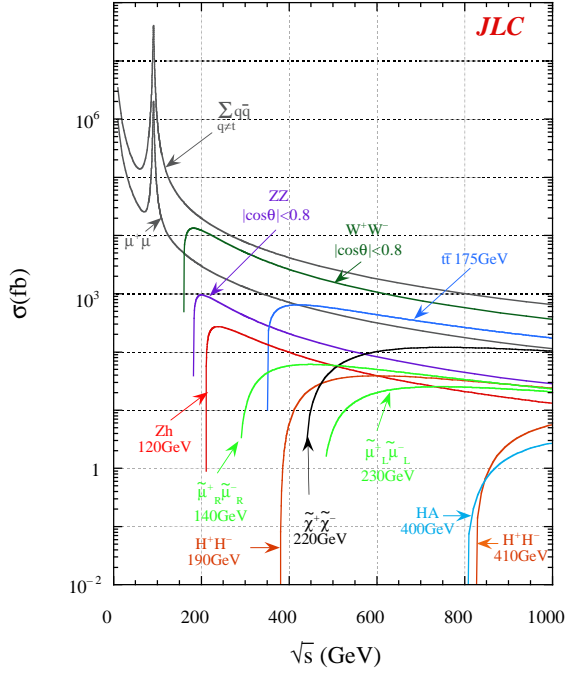}
\end{center}
\caption{Production cross sections for some standard model particles as well as
for new physics particles at $e^+e^-$ collider as functions of c.m. energy. \cite{JLCphys}}
\label{fg:prodcross}
\end{figure}

We start from the particles that are ingredients
of the standard model. Their properties and interactions with
other particles, 
however, may reveal physics beyond the standard model.
The goal is to look at the behavior of the members of
the standard model to see if there is any hint of new physics.
Production cross sections for some standard model particles as well as
those for particles beyond the standard model are shown in 
Figure \ref{fg:prodcross}.

\subsection{Gauge Bosons}

Non-Abelian nature of gauge group leads to couplings among
gauge bosons, and their pattern reflects the structure of
the underlying gauge group. $W$-pair creation $e^+e^-\to W^+W^-$
is highly sensitive to the triple gauge couplikngs $WW\gamma$ and 
$WWZ$ which can be separated by beam polarizations. With 90\% and 60\%
for electron and positron polarizations, respectively, and
500 fb$^{-1}$ at $\sqrt{s}=$500 GeV and 1 ab$^{-1}$ at $\sqrt{s}=$800 GeV,
anomalous couplings can be measured with typical 
errors of $10^{-3}$ relative.\cite{GB3} The $WW\gamma$ magnetic dipole
coupling $\kappa_\gamma$, in particular, can be measured to $10^{-4}$,
which is more than order of magnitude better than LHC with the
same years of running. The triple gauge coupling $WW\gamma$ can also be
studied by the single gauge boson productions 
$e^+e^-\to e^-\nu W^+, \nu\nu Z$, and also
by the $e\gamma$ and $\gamma\gamma$ options; namely,
$e^-\gamma\to W^-\gamma$ and $\gamma\gamma\to W^+W^-$ where the $WWZ$
coupling does not contribute.

If Higgs is not found at LHC or ILC, it may indicate that $W$-pair
can form a bound state which could be found in the $WW$ scattering
process $e^-e^+\to \nu\bar\nu W^+W^-$ as a resonance or anomalous
quartic gauge couplings. Quartic gauge couplings can generally be probed by
gauge boson scattering processes of the type $e^-e^+\to VVf\bar f$
where $V$ is $W$or $Z$ and $f$ is $e$ or $\nu$, or by triple
gauge boson productions $e^-e^+\to VVV$. At ILC, one can tell the
initial and final states of the gauge boson
scatterings, which is often difficult at LHC.

If no Higgs or no new particles are found, precision measurements on $Z$
become important. The Giga-$Z$ option can collect 1 billion $Z$'s in a few
months, and can improve by more than one order of magnitude those measurements
that use $b$-tagging and/or beam polarizations.\cite{Giga-Z} The improved $b$-tagging
is realized by the excellent vertexing capability of ILC detectors.

Couplings of fermions and gauge boson can also be studied by $e^+e^-\to f\bar f$
($f$ stands for a fermion),
where anomalous couplings may be parametrized by
$(1/\Lambda_{ij}^2)(\bar e_i\gamma^\mu e_i)(\bar f_j\gamma_\mu f_j)$ $(ij=L,R)$.
ILC is sensitive to $\Lambda_{ij}$ of typically 20 to 100 TeV.\cite{eeff}

The $e^+e^-\to f\bar f$ modes are also sensitive to existence of 
an extra $Z$ boson ($Z'$) even when the mass of $Z'$ is above the CM energy. 
Such extra gauge bosons appear in many extensions of the standard model.
Some examples are the $E_6$ $\chi$ model ($\chi$), left-right symmetric model (LR),
Littlest Higgs model (LH), Simplest Little Higgs model (SLH), and model with extra dimensions
where $Z'$ particles are actually spin-2 Kaluza-Klein excitations of gravitons (KK).
The signatures appear in
the forward-backward asymmetry of the $f\bar f$ production and 
in the dependence of the
cross section on the beam polarization. 
\begin{figure}[tb]
\begin{center}
\includegraphics{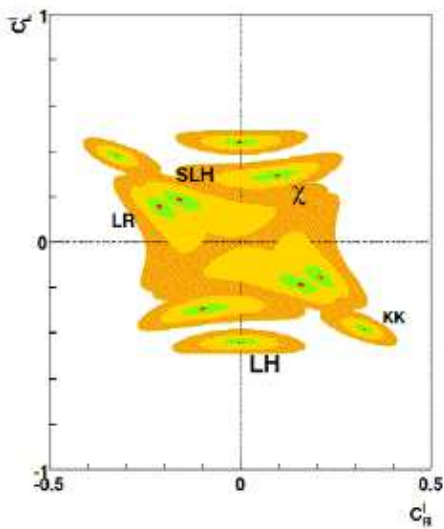}
\end{center}
\caption{The 95\%\ confidence level regions for various models using
$e^+e^-\to$ lepton pair.\cite{Zprime} The regions correspond to $m_{Z'}$ = 1,2,3,4 TeV
with the smallest region is for 1 TeV. }
\label{fg:Zprime}
\end{figure}
The resolving power of ILC in the 2-dimensional space of $C^\ell_L$ and $C^\ell_R$
is shown in Figure \ref{fg:Zprime} for $e^+e^-\to \mu^+\mu^-$, where $C^\ell_{L,R}$ are
the left-handed and right-handed $Z'\ell$ coupling coefficients where the lepton universality
is assumed. Electron and positron polarizations of 80\%\ and 60\%\, respectively, are assumed.
There are quadratic ambiguities due to the sign-independence of coupling coefficients.
LHC may find a $Z'$ resonance, but it would take ILC to identify the underlying theory.

\subsection{Top quark}

The top quark is the heaviest elementary particle observed so far, and its mass
$\sim$174 GeV is in the range of the electroweak symmetry breaking. Its large mass
indicates that it couples to Higgs strongly and thus should be 
sensitive to the structures
in the Higgs sector, or whatever is responsible for creation of masses.
In many models beyond the standard model, the Higgs mass strongly depends on
the top mass. In MSSM (the minimal supersymmetric standard model), for example, an
error in the top mass corresponds to a similar error in the Higgs mass,
which means that precision measurements of the top and Higgs masses serve
as a stringent test of theoretical models. In some cases, non-standard
top couplings may be the only area new physics can be found.
\begin{figure}[tb]
\begin{center}
\includegraphics{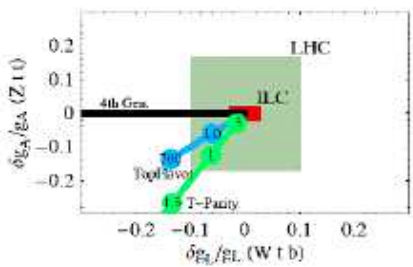}
\end{center}
\caption{Sensitivities of ILC and LHC on the axial $ttZ$ coupling and the 
left-handed $tbW$ coupling.\cite{tV}
The expected deviations for the top-flavor model and the little-Higgs
model with T-parity, and the model with 4th generation are also shown.}
\label{fg:tV}
\end{figure}

The top mass $m_t$ is best measured by the $e^+e^-\to t\bar t$ threshold scan, taking
about 5 fb$^{-1}$ each at several points of CM energy. 
Since the top quark decays before it hadronizes, the excitation curve, i.e. the cross
section as a function of CM energy,
around the threshold can reliably be calculated. It is affected by the beam energy spread,
initial-state radiation, beamstrahlung (radiation from a beam particle under the coherent
electromagnetic field of the incoming bunch), as well as the higher order corrections
which has been performed up to including some of the next-to-next-to-leading 
logarithms (NNLL).\cite{topNNLL}
The experimental and theoretical uncertainties are of the same order, and the resulting
overall error on $m_t$ is expected to be 100 to 200 MeV which can be compared to
1 to 2 GeV at LHC. The threshold scan also yield the top width to a few \%\ of
its value which is around 1.5 GeV. 

The production and decay of top quark in $e^+e^-\to t\bar t$, $t\to bW$ can be
studied near the threshold, well above the threshold, or below the threshold
(where one of the top quark is off-shell). The production is sensitive to
$ttZ$ and $tt\gamma$ couplings and the decay is sensitive to $tbW$ coupling. 
Many beyond-the-standard models predict deviations in these couplings from the
standard-model values. 
The models with 4th generation with large mixing between 4th and 3rd generations of
quarks would have the $tbW$ coupling smaller than that of the standard model while
the $ttZ$ coupling would be the same. The little Higgs models with T-parity and
the top flavor models would have both $tbW$ and $ttZ$ couplings smaller than
those of the standard model. Figure \ref{fg:tV} shows the sensitivities of ILC and LHC on
the axial $ttZ$ coupling and the left-handed $tbW$ coupling as well as
the expected deviations for the top-flavor model and the little-Higgs
model with T-parity, and the model with 4th generation. The numbers shown on the 
line for T-parity are the strength of the Higgs-top-(top partner) coupling and those
on the line for the top flavor model are the mass of the extra $Z$ boson.
Furthermore, the Kaluza-Klein mode
of graviton with mass 10 to 100 TeV in Randall-Sundrum models\cite{RS} 
may be indirectly detected as anomalous $t\bar t$ production. 
 
\subsection{Higgs particle}

Our current knowledge on the mass of the Higgs particle mainly comes from
the LEP experiments.\cite{mHlim} Within the framework of the standard model,
Higgs mass $m_H$ is bounded as $114.4<m_H<166$ GeV at 95\%\ confidence level,
where the lower limit is from direct searches and the upper limit is by an overall
fit of the standard model parameters to the data. On the other hand, 
Higgs in the MSSM is constrained to be less than 135 GeV, 
which is lower than the upper limit in the standard model.
These Higgs particle, if they exist, will be found at LHC within the first few years of
running. At ILC, even though the start would be many years later than LHC, the same
level of discovery sensitivity can be obtained by one day of running at the design luminosity.
With its clean initial and final states, and high resolutions of the ILC detectors, ILC
will be able to perform measurements on spin and parity of the Higgs particle, and
determine coupling strengths to various particles model-independent ways.

\begin{figure}[tb]
\begin{center}
\includegraphics{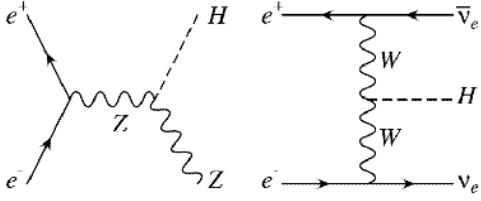}
\end{center}
\caption{The main Higgs production mechanisms at ILC:
the Higgs-strahlung (left) and the $WW$ fusion (right).}
\label{fg:Hprod}
\end{figure}
\begin{figure}[tb]
\begin{center}
\includegraphics[width=3in]{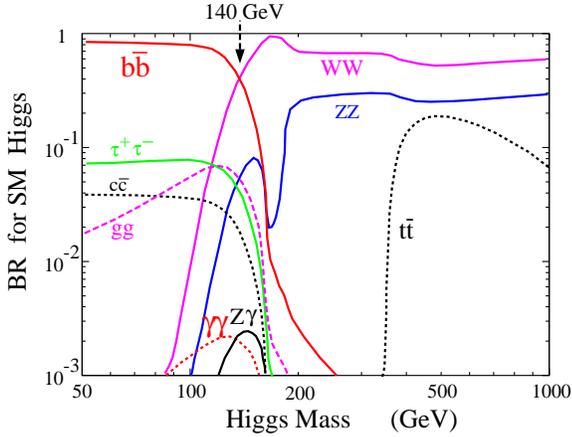}
\end{center}
\caption{Higgs decay branching fractions as functions of Higgs mass.\cite{JLCphys}}
\label{fg:higgsbrg}
\end{figure}

The primary production channels of the standard model Higgs are
$e^+e^- \to Z^* \to ZH$ (Higgs-strahlung) and $e^+e^- \to \nu\bar\nu H$
($WW$ fusion) as shown in Figure \ref{fg:Hprod}.
The Higgs-strahlung dominates at low CM energies ($<$500 GeV) and the
$WW$ fusion dominates at high CM energies ($\sim 1$ TeV). For $m_H$ of 120 GeV,
an integrated luminosity of 500 fb$^{-1}$ at CM energy of 500 GeV will
generate $3\sim 4\times10^{4}$ Higgs particles in each of the two production channels.
The decay branching fractions of Higgs are shown in Figure \ref{fg:higgsbrg}.
If the Higggs mass is below around 140 GeV, it decays primarily to $b\bar b$ with
a few \%\ each for $c\bar c$, $\tau\bar\tau$ and $gg$ branching franctions. 
The width of Higgs in this mass range is less than 10 MeV.
For $m_H$ larger than
around 150 GeV, it decays primarily to $WW$ with the $ZZ$ channel following at 20\%\ level.
The $t\bar t$ final state opens for $m_H$ larger than around 350 GeV and peaks
for $m_H\sim 500$GeV at 20\%\ branching fraction. At $m_H$ of around 500 GeV,
the Higgs is quite broad with $\Gamma_H\sim 100$GeV.

\begin{figure}[tb]
\begin{center}
\includegraphics[width=3in]{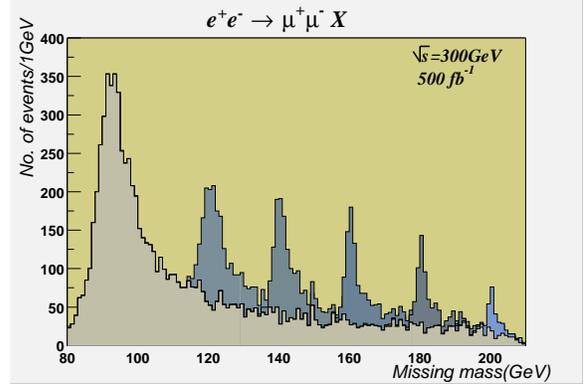}
\end{center}
\caption{The $\mu\mu$ recoil mass distribution in $e^+e^-\to \mu\mu X$
with 500 fb$^{-1}$ at CM energy of 300 GeV.\cite{JLCphys} The peaks correspond to 
$e^+e^-\to ZH$ followed by $Z\to\mu\mu$ with different values of
$m_H$, together with the background from $e^+e^-\to ZZ$, $Z\to\mu\mu$.\cite{GLCroadmap}
}
\label{fg:ZHmumu}
\end{figure}

Figure \ref{fg:ZHmumu} shows the recoil mass distribution for $e^+e^-\to ZH$,
$Z\to\mu\mu$ with 500 fb$^{-1}$ at CM energy of 300 GeV. 
Peaks corresponding to different values of
$m_H$ are shown together with the background from $e^+e^-\to ZZ$ followed by one or both of the
$Z$'s decaying to $\mu\mu$. Since the Higgs particle is not reconstructed, the method
is independent of the Higgs decay modes including the case where the decay is
invisible. The range of detectable Higgs mass reaches 
close to the CM energy itself; more precisely, up to CM energy minus $m_Z$.

The Higgs mass is obtained from the recoil masss distribution itself.
Under the same conditions used for Figure \ref{fg:ZHmumu}, the error in $m_H$ is
$\sim$70 MeV which improves to $\sim$40 MeV if hadronic decays of $Z$ are
included. 
The spin and parity of the Higgs particle can be determined by 
the threshold excitation curve and the angular distribution of the Higgs
production in the Higgs-strahlung process. If the rise of the cross
section just above the threshold is $\sigma\propto\beta_H$, the $ZH$ pair
is in a S-wave. Then the parity conservation in 
$Z^*\to ZH$ indicates that the parity of Higgs is plus. At well above
threshold, $Z$ in the final state is mostly helicity 0. Since the
intermediate $Z^*$ is polarized along the beam direction, the
angular distribution of spin-0 Higgs is given by 
$|d^1_{1,0}(\theta)|^2\propto\sin^2\theta$.
The spin parity of Higgs can also be checked in
$e^+e^-\to ZH\to f\bar f f\bar f$ or in $H\to WW^*, ZZ^*\to  f\bar f f\bar f$
where $f$ stands for a fermion.\cite{H4f}
One can also study the spin correlation of the final state $\tau$'s in 
$H\to\tau^+\tau^-$ to extract the $CP$ of Higgs.\cite{HtauCP}

The Higgs-strahlung process allows one to measure the $ZZH$ coupling independently
of the Higgs decay modes. On the other hand, the $WW$ fusion process gives the
$WWH$ coupling. At low CM energy, the $WW$ fusion process $e^+e^-\to \nu\bar\nu H$
has a substantial background coming from $e^+e^-\to ZH$, $Z\to\nu\bar\nu$ which 
can be removed by looking at the recoil mass of Higgs. Also, the $WW$ fusion process
can be turned off and on by switching the beam polarizations to identify
the contribution from the $WW$ fusion process. 
The $WWH$ coupling can also be extracted from the $H\to WW^*$ branching
fraction.
The statistical errors on
$WWH$ and $ZZH$ couplings for $m_H$ of 120 GeV are $1\sim2$\%.
For the Higgs mass below 150 GeV,
the couplings of Higgs to $b$, $c$, and $\tau$ are measured by
reconstructing the Higgs decays to $b\bar b$, $c\bar c$, and $\tau^+\tau^-$ 
in the Higgs-strahlung process. Here, the branching ratios are 
proportional to the square of the fermion mass, and the excellent
vertexing capability of ILC detectors is essential in separating $c\bar c$
from $b\bar b$. The $t\bar tH$ Yukawa coupling is measured by
$e^+e^-\to t \bar t^* \to t \bar t H$ at ~ 1 TeV.
The process $e^+e^-\to t \bar t$ is itself
sensitive to the $t\bar tH$ coupling through the $H$-loop vertex correction. 
The gluonic decay $H\to gg$ as well as the
decays $H\to \gamma\gamma, \gamma Z$ are 
sensitive to the $t\bar tH$ coupling though top loop, and also
sensitive to new heavy particles that may contribute in the loop.
For high Higgs masses, the gauge boson pair final states dominate.
Still, with 1 ab$^{-1}$ at 1 TeV, the $b\bar b$ branching fraction can be measured to 
12\%\ and 28\%\ for $m_H=$ 180 and 220 GeV, respectively. 
Invisible final state can also be found by the recoil mass technique,
with $5\sigma$ confidence down to 2\%\ 
branching fraction for $120 < m_H < 160$ GeV.

The total Higgs width for $m_H$ less than $\sim$200 GeV is too narrow
to be measured directly, but can be indirectly measured by 
$\Gamma_H = \Gamma(H\to WW^*)/Br(H\to WW^*)$ where $Br(H\to WW^*)$ is
directly measured and $ \Gamma(H\to WW^*)$ is estimated from the measurement
of the $WWH$ coupling by, say, the $WW$ fusion process. For $120 < m_H < 160$ GeV,
the total Higgs width can be measured with an error of 4 to 13\%.

The trilinear Higgs coupling, or the Higgs self coupling, can be
measured by $e^+e^-\to ZH^*\to ZHH$ or by 
$e^+e^-\to \nu\bar\nu H^*\to \nu\bar\nu HH$. 
The cross section is quite small and the final state
$(b\bar b) (b\bar b)(\ell^+\ell^-)$ challenges the capability
of detector. Here, a superb vertexing resolution is critical for
the $b$ tagging, and an excellent jet energy reconstruction is
needed for calculating the invariant masses of $b$ jet pairs.
With 1 ab$^{-1}$ at 500 GeV CM energy and for $m_H =$ 200 GeV,
the error on the Higgs self coupling constant $\lambda_{HHH}$ is estimated to be
about 20\% using the $e^+e^-\to ZH^*\to ZHH$ mode only.\cite{HHH}
If one combines $e^+e^-\to ZHH$ and $e^+e^-\to \nu\bar\nu HH$
at 1 TeV CM energy, the error in $\lambda_{HHH}$ becomes
12\%\ for the same Higgs mass with 1 ab$^{-1}$ and 80\%\
electron polarization.\cite{Yasui}

\begin{figure}[tb]
\begin{center}
\includegraphics{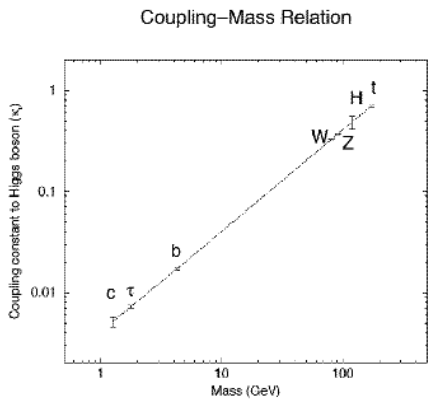}
\end{center}
\caption{The Higgs coupling constants as functions of mass of the particle that
Higgs couples to.\cite{GLCroadmap} The couplings with gauge bosons and the self coupling
are normalized differently from those with fermions.}
\label{fg:Hcoups}
\end{figure}

Expected results for Higgs coupling measurements are plotted in Figure \ref{fg:Hcoups}
as functions of mass of the particle that Higgs couples to.
Coupling constants of Higgs to fermions, weak bosons $W$ and $Z$, and 
Higgs itself are given by $m_f/a$, $g\, m_W$, $g \, m_Z/2\cos\theta_W$, and
$m_H^2/2a$, respectively, where $g\sim0.65$ is the $SU(2)$ coupling constant, 
and $a\sim246$ GeV is the vacuum expectation value of Higgs.
Thus, when properly normalized, the Higgs couplings of the standard model 
should be proportional to the mass of the particle it couples to.
The pattern of deviation from the standard model serves as
a powerful probe of the mechanism of mass generation.
For example, for a two-Higgs-doublet model where up-type fermion masses
are generated by one doublet and down-type fermion masses by another 
(so-called Type-II two-Higgs-doublet models),
the Higgs couplings to all the up-type fermiosns are shifted by a factor,
and those to all the down-type fermions are shifted by another factor.
And in models with Radion-Higgs mixing, the Higgs couplings may be
reduced uniformly with respect to the standard model values.

\section{New Physics particles}

Among the extensions of the standard model, the SUSY models occupy
a special place due to their theoretical virtues, the primary one
of which is to make the Higgs mass stable in the weak scale.
There are also other models that address the same problem, and
these models usually contain particles that do not appear in the
standard model. One should keep in mind, however, that Nature may
have in store for us something that have nothing to do with any of these,
and we may be lucky enough to encounter them at LHC/ILC.

\subsection{SUSY particles}

The minimal supersymmetric standard model (MSSM)
is the most economical model with $R$-parity
conservation which makes the lightest superparticle (LSP) stable. The LSP thus
becomes a candidate for the dark matter. The two complex Higgs doublets 
and the four massless gauge bosons have 8 charged degrees of freedom and
8 neutral degrees of freedom. After breaking of SUSY and gauge symmetries, 
their super partners mix to form
two charginos $\chi^\pm_{1,2}$ (8 degrees of freedom) and four
neutralinos $\chi_{1,2,3,4}^0$ (8 degrees of freedom) all with spin 1/2.
The neutralinos are self-conjugate; namely, they are Majorana particles.
For each fermion $f$, there are two spin-0 superpartners corresponding to
two helicities of the fermion: $\tilde f_{R}$ and $\tilde f_{L}$ which
could in general mix, particularly for the third generation fermions.
Since the actual masses of each particle and its super-partner are clearly
different, the supersymmetry is broken by some mechanism. One popular
model is a minimal model with gravity-mediated SUSY breaking (mSUGRA) in which
there are only four free parameters and a sign, which may be taken as 
the mass parameters of scalers and winos: $m_0$ and $M_2$, the trilinear Higgs
coupling $A_0$, the ratio of vacuum expectation values of the two Higgs
doublets, $\tan\beta$, and sign$(\mu)$ where $\mu$ is a Higgs mass parameter.
For concreteness, we look for SUSY particles 
in this section with mSUGRA as a guide. 

In many scenarios of SUSY, the super-partners of leptons (sleptons) are
light enough to be produced at ILC. In addition, they tend to decay to
the corresponding lepton plus the LSP neutralino. 
In the scenario called SPS1a of mSUGRA, 
all sleptons decay dominantly as $\tilde\ell\to \ell\chi_1^0$,
where $\ell$ is a lepton and $\tilde\ell$ is its super-partner.
Decays and interactions of right-handed sleptons are particularly
simple since they are $SU(2)_L$ singlets and thus do not interact 
with $SU(2)_L$ gauge particles.
\begin{figure}[tb]
\begin{center}
\includegraphics{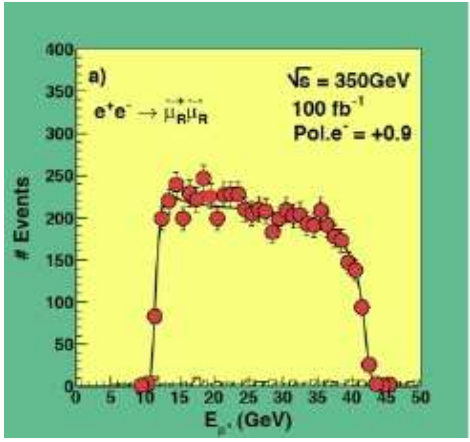}
\end{center}
\caption{The muon energy distribution in smuon pair production at well above
 threshold,\cite{GLCroadmap}
 $e^+e^-\to \tilde\mu_R^+ \tilde\mu_R^-$,  $\tilde\mu_R^+ \to \mu^+ \chi_1^0$.
 The high and low end points gives both $m_{\chi}$ and $m_{\tilde\mu}$.}
\label{fg:smuon}
\end{figure}
Figure \ref{fg:smuon} demonstrates simultaneous mass determination of 
the right-handed smuon $\tilde\mu_R$ and the LSP neutralino $\chi_1^0$ in 
$e^+e^-\to \tilde\mu_R^+ \tilde\mu_R^-$ followed by 
$\tilde\mu_R^+ \to \mu^+ \chi_1^0$ and its charge conjugate mode.
The data is taken well above the threshold with 100 fb$^{-1}$ at 
350 GeV CM energy. The smuon and the LSP masses are assumed to be 142 GeV and
118 GeV, respectively. 
The high and low end points of the muon energy distribution gives
both masses to a few $\times10^{-3}$ of themselves. 
This is in contract to the LHC case where the mass of
LSP is difficult to measure directly. This mode also illustrates
the effectiveness of beam polarization in background reduction.
\begin{figure}[tb]
\begin{center}
\includegraphics{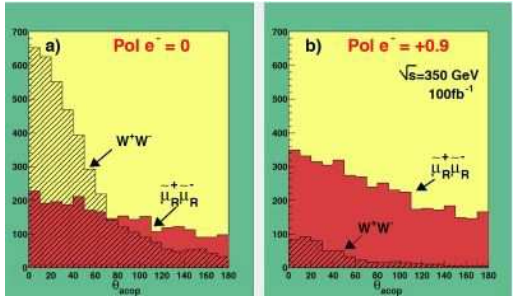}
\end{center}
\caption{The muon acoplanarity distributions for 
   the smuon production\cite{GLCroadmap} $e^+e^-\to \tilde\mu_R^+ \tilde\mu_R^-$, 
   $\tilde\mu_R^\pm \to \mu^\pm \chi_1^0$. With no electron polarization (left)
   and with 90\%\ electron right-handed polarization.}
\label{fg:smuopol}
\end{figure}
The muon acoplanarity distribution in $e^+e^-\to \tilde\mu_R^+ \tilde\mu_R^-$, 
$\tilde\mu_R^+ \to \mu^+ \chi_1^0$ is shown in Figure \ref{fg:smuopol}
for no electron polarization and with 90\%\ electron polarization.
Here, the acoplanarity angle is the angle between the muon pair projected to
a plane perpendicular to the beam line.
By polarizing the electron right-handedly, one can eliminate the
background caused by $e^+e^-\to W^+W^-$, $W^+\to\mu^+\nu$ and its
charge conjugate. This is because for the s-channel the initial state
$e^-_R$ limits the intermediate state to $B$ (the gauge boson of hypercharge Y)
which does not couple to $W$ in the final state, and the t-channel
neutrino exchange is a $V-A$ interaction which does not couple to $e^-_R$.

The angular distribution of the smuon production 
should be $\sin^2\theta$ since smuon is spin 0 and the intermediate $Z/\gamma$
state is polarized as $|1,\pm1 >$ along the beam line since the
electron coupling to the intermediate state is a linear combination of
vector and axial vector. The production
angle can be reconstructed with a quadratic ambiguity where the
wrong solution has a flat distribution that can be subtracted.
The resulting angular distribution can be checked to
be consistent with the expected shape. 

The smuon mass can also be
determined at the threshold, where an energy scan gives the
threshold excitation curve which should rise slowly as $\beta_{\tilde\mu}^3$
due to the $P$ wave nature of the smuon pair. 

A large mixing effect is expected for the stau sector and $\tilde\tau_R$ and 
$\tilde\tau_L$ would mix to form mass eigenstates $\tilde\tau_1$ and 
$\tilde\tau_2$ where $\tilde\tau_1$ is defined to be the lighter of the two.
The mixing angle can be determined by two or more measurements of
$e^+e^- \to \tilde\tau_1^+\tilde\tau_1^-$ with different beam polarizations.
In the SPS1a scenario mentioned earlier, $\tilde\tau_1$ is the lightest of the
sleptons with its mass around 100 GeV, and the dominant decay
is $\tilde\tau_1 \to \tau \chi_1^0$. In this case, the mixing angle 
($\cos2\theta$) can be determined at the percent level. 

The situation for the chargino pair production is similar to that of
smuon pair production: 
$e^+e^-\to \chi^+_1 \chi^-_1$ followed by $\chi^\pm_1\to \chi^0_1 W^\pm$,
where the energy distribution of $W^\pm$ simultaneously determines
the masses of the chargino $\chi_1^\pm$ and the LSP neutralino.
With the mass of the LSP obtained in the smuon study, the mass of
the lightest chargino $\chi_1^\pm$ can be determined to 1 \%\ level.

For neutralinos, the invariant mass distribution of the lepton pair in 
$e^+e^-\to \chi^0_2 \chi^0_1$ followed by $\chi^0_2\to \chi^0_1 \ell^+\ell^-$
can determine the mass difference between $ \chi^0_2 $ and $\chi^0_1$
to better than 1\%. This mode may also demonstrate a sizable CP violation
effects for some parameter space of MSSM. For example, 
the sign asymmetry of the $T$-odd triple 
product $p_{e^-}\cdot (p_{\ell^+}\times p_{\ell^-})$ can be as large as
20\%\cite{Hesselbach}. Similar $T$-odd triple products can be
formed for other modes such as $e^+e^-\to \chi^+_1 \chi^-_1$.

The ability to select the beam polarization allows us to
probe into the structures of the SUSY models. For example,
the charginos $\chi_{1,2}^\pm$ are the mass eigen states of the system composed of 
the charged Higgsinos ($\tilde H_u^+$, $\tilde H_d^-$) 
and charged gauginos $(\tilde W^\pm)$ where the
mass matrix term can be written as
\[
\begin{pmatrix} \tilde W^+ & \tilde H_u^+ \end{pmatrix}
\begin{pmatrix} M_2 & \sqrt2 m_W \cos\beta \cr
        \sqrt2 m_W \sin\beta & \mu \end{pmatrix}
\begin{pmatrix} \tilde W^- \cr \tilde H_d^- \end{pmatrix}\;.
\]
By using a right-handed electron beam for $e^+e_R^-\to \chi^+_1 \chi^-_1$,
the intermediate s-channel state is purely $B$ which is the gauge boson
for hypercharge $Y$.
On the other hand, $B$ couples only to the Higgsino component of 
chargino; thus, 
one can obtain information on the mixing parameters of the charginos.
Together with cross section measurements of $e^+e_R^-\to \tilde e^+_R \tilde e^-_R$ which is
sensitive to the mass parameter $M_1$ which is the
mass parameter for Bino (superpartner of $B$),
one can perform a global fit to the
parameters $(M_1, M_2, \mu, \tan\beta)$. 
\begin{figure}[tb]
\begin{center}
\includegraphics{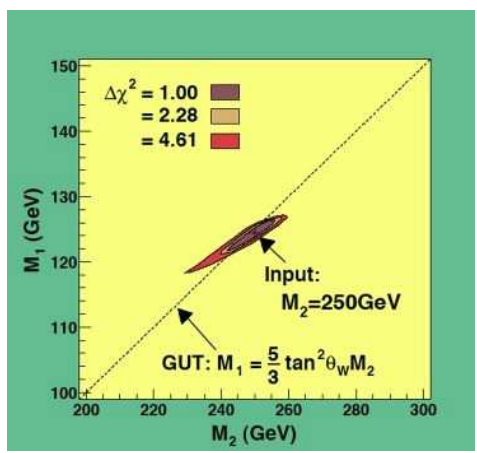}
\end{center}
\caption{The result of a global fit\cite{GLCroadmap} to 
 $e^+e_R^- \to \tilde e_R^+ \tilde e_R^-$ and $e^+e_R^-\to \chi^+_1 \chi^-_1$.
 The line shows the GUT relation between $M_1$ and $M_2$.}
\label{fg:M12GUT}
\end{figure}
Figure \ref{fg:M12GUT} shows the result of the global fit.
If the masses $M_1$ and $M_2$ are to converge to a single value at the
GUT scale, they would satisfy the GUT relation
\[
  M_1 = {5\over 3} \tan^2\theta_W M_2\;,
\]
which is tested in a highly model independent way.

In most SUSY scenarios, squarks are in general heavier than sleptons and 
many of them are beyond the reach of ILC even with the energy upgrade to 1 TeV.
Still, due to the large mixing effects expected for the third generation squarks
$\tilde t$ and $\tilde b$, the lighter ones, $\tilde t_1$ and $\tilde b_1$,
can be within reach of ILC. When they can be pair produced as in
$e^+e^- \to \tilde t_1 \tilde t_1$, then multiple measurements of cross sections
with different beam polarizations can determine the mixing angle just as in
the case of the stau pair production with a similar precision. 

\subsection{Kaluza-Klein (KK) mode gravitons}

In the models with large extra dimensions where only gravitons can propagate 
in the extra dimensions, the fundamental gravity mass
scale $M_D$ can be as small as the TeV scale.\cite{extradim}
When the wave function of the graviton has a certain number of nodes
in the direction of the extra dimensions (Kaluza-Klein modes),
it can have mass as a function of the number of nodes.
When the number of the extra dimension $\delta$ is 2 to 6, the
size of extra dimension can be very large and is around 0.1 mm to 1 fm,
for which the KK mode graviton $G_{KK}$ has effectively a continuous mass spectrum.
At ILC, one may search for emission of KK mode graviton in
$e^+e^-\to \gamma G_{KK}$ where $G_{KK}$ escapes the detector
and appear as missing energy. Here again, the beam polarization is
a powerful handle to suppress the main background $e^+e^-\to \nu\bar\nu\gamma$.
With 1 ab$^{-1}$ at 800 GeV and 
with the electron and positron beam polarizations of 80\%\ and 60\%\, respectively,
the 95\%\ confidence level lower limit of $M_D$ is 10 (3) TeV for $\delta$ of
2 (6). This is similar to the sensitivities at LHC. At ILC, however, 
one can utilize the angular distribution of $\gamma$ to verify the
spin of $G_{KK}$ which should be two. In addition, 
the number of extra dimension $\delta$ can be measured at ILC by the energy
dependence of the cross section, say at 500 GeV vs at 800 GeV, 
and the missing mass distribution.

\subsection{Little Higgs models}

In the Little Higgs models, the Higgs particle is composite, and there exist
extra gauge bosons and top partners.\cite{littleH} 
Most new particles are too heavy to be
directly detected at ILC, but indirect search for extra gauge bosons 
is possible with $e^+e^- \to f\bar f$ as described earlier (Figure \ref{fg:Zprime}).
Furthermore, in the model with $T$-parity, there could be
a pseudo-axion $\eta$ below 1 TeV. In such cases, $e^+e^-\to ZHH$ can be
substantially enhanced by $ZH\eta$ coupling: $e^+e^-\to Z^* \to \eta^*H$, $\eta^*\to ZH$
which should be easily detectable with
the TeV upgrade of the machine.

\subsection{Cosmological connections}

\begin{figure}[tb]
\begin{center}
\includegraphics{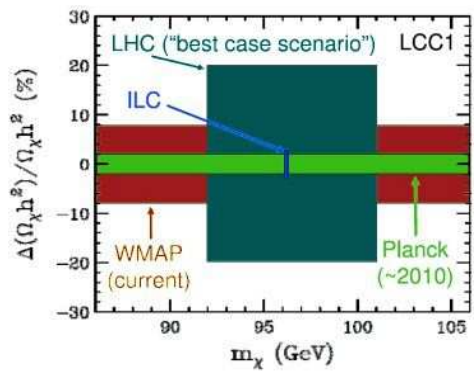}
\end{center}
\caption{The error in $\Omega_{DM}$ vs the lightest neutralino mass
as constrained by LHC and ILC data within the framework of
mSUGRA with parameter set SPS1a.\cite{cosmo} }
\label{fg:cosmo}
\end{figure}

The WMAP satellite data indicates that the cold dark matter density
of the universe is given by $\Omega_{DM} h^2 = 0.113\pm0.009$ and
makes up about 1/4 of the energy of the universe.\cite{WMAP}
The error on $\Omega_{DM}$ will be reduced significantly by 
the Planck measurements expected around 2010. 
In the MSSM, the 
lightest neutralino $\chi^0_1$ serves as a candidate for the cold dark matter.
In order to predict the relic density of the cold dark matter, however,
all interactions contributing to $\chi_1^0$ annihilation should be known.
Figure \ref{fg:cosmo} shows the result of a study within the
mSGURA SPS1a scenario. The sensitivities of LHC and ILC
in the 2-dimensional space of $m_{\chi_1^0}$ and the 
estimated error on $\Omega_{DM}$ are shown together with the uncertainties
on $\Omega_{DM}$ by WMAP and Planck. 
ILC can determine the mass of $\chi_1^0$ much more accurately than LHC, and
the error on the estimate of $\Omega_{DM}$ is comparable to the
error expected for the future measurement by Planck.

\section{$Options: \gamma\gamma$ and $e^-e^-$ colliders}

The $e^+e^-$ mode of ILC can accomodate $e^-e^-$ and
$\gamma\gamma$ colliders with relatively minor modifications.
The $\gamma\gamma$ collider requires a pair of powerful lasers
that are aimed at the interaction point from both sides along the
beam line. The photons that are Compton back-scattered
by incoming beams collide at the interaction point. The maximum CM energy 
of the $\gamma\gamma$ collision is only slightly lower than that 
of the $e^+e^-$ collision, and the luminosity is also comparable.
The disrupted beams, however, need to be extracted without hitting the
sensitive detector parts, and this necessitates a crossing angle greater
than 25 mrad (compared to the nominal 14 mrad). Also, the original
beams also collide on top of the $\gamma\gamma$ collisions, 
and this favors the $e^-e^-$ mode over the $e^+e^-$ mode which has larger
total cross section. The $e^-e^-$ is suited for the $\gamma\gamma$ collision
also because it is easier to produce polarized 
electrons than polarized positrons. 

The Higgs particle can be produced by the s-channel $\gamma\gamma \to H$
process which involves loop diagrams of charged particles.
It allows a precision measurement of the Higgs coupling to
photon. and is sensitive to new particles that can 
contribute in the loop.
The Higgs mass reach is close to the CM energy of the $e^-e^-$ beams itself. 
Higgs below 140 GeV would be detected in the $b\bar b$ final state.
With 410 fb of $\gamma\gamma$ luminosity at the beam CM energy of 210 GeV, and 
for $m_H =$ 120 GeV, $\Gamma(H\to \gamma\gamma)$$\times$$Br(H\to b\bar b)$
can be determined to a statistical error of 2 \%.\cite{ggHiggs}
Even for heavy Higgs of 200 to 350 GeV, the two photon width
can be determined with errors of 3 to 10\%. The total Higgs decay
width can be obtained by combining the $\Gamma(H\to \gamma\gamma)$ measurement
with $Br(H\to \gamma\gamma)$ measured at the $e^+e^-$ collider at high
CM energy. The expected error on the Higgs total width is about 5\%\ 
for $m_H =$ 120 to 140 GeV, which 
is competitive with the method using 
$\Gamma_H = \Gamma(H\to WW^*)/Br(H\to WW^*)$ mentioned earlier.

The $e^-e^-$ collider can generate exotic charge 2 particles in s-channel.
It is also sensitive to Majorana neutrino exchange in $e^-e^- \to W^-W^-$. 
The neutralino exchange interactions $e^-e^- \to e_{L,R}^-e_{L,R}^-$
allows one to study the quantum numbers of selectrons through
beam polarizations.

\section{Summary}

The clean environment and the well-defined initial state of $e^+e^-$ collision,
including the spin states, as well as the superb resolutions of the ILC detectors
make the ILC physics program very attractive. ILC can study the particles 
found at LHC in detail to uncover the underlying theoretical structures, and
in some cases discover new particles and reactions that are buried in backgrounds
at LHC.

\section*{Acknowledgment}

The author would like to thank the editorial panel of the detector concept report (DCR)
who has put together an excellent summary of the ILC physics studies performed so far,
and all those who have contributed to these studies. He would like to thank, in particular,
Prof. Komamiya who has entrusted him to write this article. 
This work is supported in part by
JSPS grant 18GS0202.

\end{document}